\documentstyle[12pt]{article}

\begin{document}

\begin{titlepage}
\thispagestyle{empty}
\begin{flushright} {\small McGill/94--45;} {\small UMDGR--95--047}\\
                   {\small gr--qc/9503020}\\
\end{flushright}

\vfill

\begin{center}
{\bf \huge Increase of Black Hole Entropy}\\
{\bf \huge in Higher Curvature Gravity}\\[2ex]

\vfill

{\large
Ted Jacobson$^{a,}$\footnote{E-mail address: jacobson@umdhep.umd.edu},
Gungwon Kang$^{a,}$\footnote{E-mail address: eunjoo@wam.umd.edu},
and Robert C. Myers$^{b,}$\footnote{E-mail address:
rcm@hep.physics.mcgill.ca}\\}
\vspace{.25in}
{\it $^a$Department of Physics, University
of Maryland, College Park, MD 20742--4111}\\
\vspace{.05in}
{\it $^b$Department of Physics, McGill University, Montr\'{e}al, Qu\'{e}bec,
Canada H3A 2T8}\\

\end{center}
\vfill
\begin{abstract}
{\rm
We examine the Zeroth Law and the Second Law of black hole thermodynamics
within the context of effective gravitational actions including higher
curvature interactions. We show that entropy can never decrease for
quasi-stationary processes in which a black hole accretes
positive energy matter, independent of the details of the
gravitational action. Within a class of higher curvature
theories where the Lagrangian consists of a polynomial in the Ricci
scalar, we use a conformally equivalent theory to establish that
stationary black hole solutions with a Killing horizon
satisfy the Zeroth Law, and that
the Second Law holds in general for any dynamical process.
We also introduce a new method for establishing the Second
Law based on a generalization of the area theorem, which may prove
useful for a wider class of Lagrangians. Finally, we show how one
can infer the form of the black hole entropy, at least for the
Ricci polynomial theories, by integrating the changes of mass and angular
momentum in a quasistationary accretion process.}
\end{abstract}

\vfill

\end{titlepage}
\pagebreak

\section{Introduction}

One of the primary motivations to study black hole thermodynamics is to
gain some insight into the nature of quantum gravity. Whatever
framework physicists eventually uncover to describe
quantum gravity,
there should be a low energy effective
action which describes the dynamics of a ``background metric field''
for sufficiently weak curvatures at sufficiently long distances. On
general grounds, one expects that this effective gravity action will consist
of the classical Einstein action plus a series of
covariant, higher-dimension interactions
({\it i.e.,} higher curvature terms, and also higher derivative terms
involving all of the physical fields) induced by quantum effects.
While such effective
actions are typically pathological when considered as fundamental, they
may also be used to define mild perturbations for
Einstein gravity coupled to conventional
matter fields. It is within this latter context of Einstein
gravity ``corrected'' by higher dimension operators that we wish to
consider modifications of black hole thermodynamics.

Naive dimensional analysis suggests that the coefficients of all higher
dimension interactions in the effective lagrangian should be dimensionless
numbers of order unity times the appropriate power of the Planck length.
Thus one might worry that the effect of the higher dimension terms
would be the same order as those of quantum fluctuations, and so
there would seem to be little point in studying modifications to
{\it classical} black hole thermodynamics from higher dimension terms.
One motivation for studying the classical problem is that it is of course
possible that this naive dimensional analysis is incorrect, just as it
would be in predicting the observed value of the cosmological constant.
So given the lack of any direct experimental evidence, it is possible that
the coefficients of some higher dimension terms are larger than
expected. Further, we would
like to know whether or not consistency with classical black hole
thermodynamics places any new restrictions on these coefficients.
Moreover, it is interesting to explore black hole thermodynamics in
generalized gravity theories in order to see whether the
thermodynamic ``analogy" is just a peculiar accident of Einstein gravity,
is a robust feature of all generally covariant theories of gravity,
or is something in between.

In this ``analogy'', any black hole should behave as a heat bath.
Quantum field theory reveals that $\kappa/(2\pi)$ is the black
hole temperature\cite{field}, independent of the details of the dynamics
of the gravity theory\cite{newfoot}. Hence this is in fact a
robust feature of black hole physics in general. An important foundation for
black hole thermodynamics is then the validity of the Zeroth Law,
namely, that the surface gravity should be constant across a stationary
event horizon. If the event horizon is a Killing horizon with a regular
bifurcation surface, it is straightforward to show that the
Zeroth Law holds\cite{raczwald}. In Einstein gravity a proof of the
Zeroth Law can also be constructed without
the assumption of a regular bifurcation surface, but with the
additional assumption that the dominant energy condition\cite{barcarhaw}
holds, {\it i.e.,} for any future pointing timelike vector $u^a$
the corresponding four momentum density $P^a=-T^{ab}u_b$
is future pointing timelike. Equivalently, for all pairs of future pointing
timelike vectors $u^a,v^a$,
\begin{equation}
T_{ab}\,u^av^b\ge0.
\label{domin}
\end{equation}
In higher curvature theories, establishing the validity of the Zeroth Law
in general remains an important open question.

Using Euclidean path integral methods\cite{euclid} it is clear that,
if the Zeroth law holds, a version of the First Law of black hole
thermodynamics still applies in these higher curvature
theories. Applying these techniques to various specific theories
and specific black hole solutions showed
that the relation equating the black hole entropy with
one quarter the surface area of the horizon no longer holds in
general\cite{failS}. However, it is now known that the entropy is
given always by a local expression evaluated at the
horizon\cite{love,wald1,visser,wald2}.

Wald\cite{wald1} established this result very generally for any
diffeomorphism invariant theory via a new
Minkowski signature derivation
of the First Law of black hole mechanics.
This law applies for variations (of the dynamical fields) around a {\it
stationary} black hole background to nearby solutions,
\begin{equation}
\frac{\kappa}{2\pi}\delta S=\delta M-\Omega^{(\alpha)}\,\delta J_{(\alpha)}
\ \ .
\label{first}
\end{equation}
Here $M$, $J_{(\alpha)}$, $\Omega^{(\alpha)}$ and $\kappa$ are the
mass, the angular momentum\cite{foot1}, the angular velocity and
the surface gravity of the black hole. Wald found that $S$ can be
expressed as a {\it local} geometric density
integrated over a space-like cross-section of
the horizon, and that it is associated with the Noether
charge of diffeomorphisms under the Killing vector field that generates
the horizon.  Eq.~(\ref{first}) then has the rather remarkable
feature that it relates variations in properties of the
black hole as measured at asymptotic infinity to a variation
of a geometric property of the horizon. Given the identification
of the temperature with $\kappa/(2\pi)$, eq.~(\ref{first}) has a natural
interpretation as the First Law of thermodynamics where $S$ is
the black hole entropy. If this $S$ is truly to play the role of entropy,
it should also satisfy the Second Law of
thermodynamics as a black hole evolves --- {\it i.e.,}
$S$ should never decrease in any dynamical processes.

For general relativity, one has the celebrated result that
the black hole entropy is given by one quarter the surface area of the
horizon, $S=A/(4 G)$\cite{area}.  In this context, the Second Law is
established by Hawking's area theorem, which states that in
any classical process involving black holes, the total
surface area of the event horizon will never decrease\cite{areathm}.
An essential ingredient in the proof of this theorem is the assumption
that the null convergence condition $R_{ab}k^ak^b\ge0$ holds for all null
vectors $k^a$. This condition is implied by
the Einstein field equation
\begin{equation}
R_{ab}-{1\over2}g_{ab}R=8\pi G\, T_{ab}
\label{eom}
\end{equation}
together with the  null energy condition
\begin{equation}
T_{ab}\,k^a k^b\ge0\qquad{\rm\ for\ any\ null\ vector\ }k^a\ \ .
\label{null}
\end{equation}
Another essential ingredient is cosmic censorship --- {\it i.e.,} it is
assumed that
naked singularities do not develop in the processes of interest.

In theories where higher curvature interactions are
included along with the Einstein Lagrangian, the equations of motion
may still be written in the form of eq.~(\ref{eom}), if the contributions
from the higher curvature interactions are included in the stress-energy
tensor. Typically, these contributions spoil the energy conditions, and so
one cannot establish an area increase theorem in such theories.
However, this is not the relevant question for black hole thermodynamics.
The relevant question is whether or not the quantity $S$
whose variation appears in the First Law (\ref{first})
satisfies a classical increase theorem. If so, one would have a
Second Law of black hole thermodynamics for these theories, further
validating the interpretation of $S$ as the black hole entropy.

In this paper, we investigate the validity of
the Zeroth Law and the classical Second Law
for higher curvature extensions of Einstein gravity.
Sect.~2 examines entropy increase in quasi-stationary processes. For
such processes, the Second Law arises directly from the First Law
for any theory of gravity, as long as
the matter stress-energy tensor satisfies the null energy condition.
Sect.~3 demonstrates that the Zeroth Law for stationary black holes and
the Second Law for arbitrary dynamical processes hold in a theory
where the gravitational Lagrangian is $R+\alpha R^2$.
These results are established by relating the higher curvature
theory by a conformal field redefinition to a more conventional
theory in which Einstein gravity is coupled to an auxiliary scalar field.
Our proofs are valid provided that $\alpha$ is positive and that
cosmic censorship holds for the conformally related metric.
Sect.~4 generalizes the previous results to a larger class of theories
where the gravitational action is a polynomial of the Ricci scalar.
In sect.~5, a proof of the Second Law is constructed directly
within the higher curvature theories without making (explicit)
use of the conformal field redefinition. This approach
follows closely the logic of Hawking's area theorem, and may provide
insight for the problem of establishing the Second Law in more general
theories. This analysis makes use of an ``extended" Raychaudhuri equation
which is also used to establish a new ``physical process"
derivation of the First Law and the form of the entropy for these
higher curvature theories analogous to that of
Wald for general relativity\cite{waldbook2}.
Sect.~6 presents a discussion of our results.

Throughout the paper, we consider only asymptotically
flat spaces, and we employ the conventions of \cite{wtext}.
We also adopt the standard convention of setting $\hbar=c=1$.
Further, motivated by the fact that many of the recent candidates
for a theory of quantum gravity are theories in higher dimensional
spacetimes, we will allow spacetime to have an arbitrary dimension,
$D\ge4$.

\section{Quasistationary Processes}

Here we demonstrate that in quasi-stationary processes,
the Second Law is a consequence of the First Law for any theory of gravity.
We wish to consider a dynamical process in which a small
amount of matter enters from a great distance and
drops into a vacuum black hole. The initial and final black hole
states are (approximately)
stationary. In those spacetime regions then, there are
Killing vector fields, $\xi^a$ and $\phi^a_{\scriptscriptstyle (\alpha)}$,
which asymptotically generate time translations and orthogonal
rotations\cite{foot1}, respectively. By a {\it quasi-stationary}
process we mean one in which the background spacetime is only slightly
perturbed by the infalling matter.
In order to establish a perturbation expansion
we introduce a small parameter  $\delta$ associated with the amplitude
of the infalling fields. We assume that the stress
energy tensor is order $\delta^2$, as are the resulting
perturbations of the metric. The vector fields, $\xi^a$ and
$\phi^a_{\scriptscriptstyle
(\alpha)}$, can then be extended through the intermediate spacetime region
where the accretion process occurs such that Killing's equation
still applies to leading order ({\it e.g.,} $\nabla^{(a}\xi^{b)}=
O(\delta^2)\,)$.
This extension is further chosen so that the horizon generator
is given by $\chi^a=\xi^a+\Omega^{\scriptscriptstyle(\alpha)}
\phi^a_{\scriptscriptstyle (\alpha)}$
throughout the entire evolution, where $\Omega^{\scriptscriptstyle(\alpha)}$ is
constant to order $\delta^2$.

Two space-like surfaces, $\Sigma_i$ and $\Sigma_f$, are introduced at
a time before the accretion process begins, and at a time after the process
has ended, respectively.
The black hole horizon $H$ provides an inner surface, and we introduce
a surface $O$ at some large radius in the asymptotically flat region.
These four surfaces enclose a spacetime
volume $V$, and we assume that the matter enters from the asymptotically
flat region and exits either as it crosses the black hole horizon
or after scattering back out across $O$. One can evaluate the mass
and angular momentum carried in with the new matter by flux
integrals in the asymptotic region,
\begin{eqnarray}
\Delta M&=&\int_OT_{ab}\,\xi^ad\Sigma^b
\nonumber\\
\Delta J_{\scriptscriptstyle (\alpha)}&=&-\int_O
T_{ab}\,\phi^a_{\scriptscriptstyle (\alpha)}d\Sigma^b,
\label{change}
\end{eqnarray}
where $d\Sigma^b$ is defined with an outward pointing normal.
In the framework of the perturbation expansion, these
quantities are $O(\delta^2)$. In what follows higher
order terms will be ignored. These expressions may be combined
to produce a flux integral containing the horizon generating vector
field $\chi^a$
\begin{equation}
\Delta M-\Omega^{\scriptscriptstyle(\alpha)}
\Delta J_{\scriptscriptstyle (\alpha)}
=\int_OT_{ab}\,\chi^ad\Sigma^b\ \ .
\label{eqone}
\end{equation}
By the First Law (\ref{first}), this combination of variations
is proportional to the variation of the entropy, ${\kappa\over2\pi}
\Delta S$. (Here, we apply the First Law in what Wald calls the
``physical process" form\cite{waldbook2}, for which one assumes
that the accretion process is a mild perturbation, and hence
that no singularities develop in $V$.)

Using stress-energy conservation as well as Killing's equation,
one has
\[
\nabla^a(T_{ab}\,\chi^b)=\nabla^aT_{ab}\,\chi^b+
T_{ab}\,\nabla^a\chi^b=O(\delta^4).
\]
Now one can integrate the above expression over $V$ and apply
Gauss' law to produce a flux integral. Since the
stress-energy vanishes on $\Sigma_i$ and $\Sigma_f$, one finds
\begin{equation}
O(\delta^4)=\int_V\nabla^a(T_{ab}\,\chi^b)\,d\Sigma=
\int_OT_{ab}\,\chi^ad\Sigma^b
+\int_HT_{ab}\,\chi^ad\Sigma^b\ \ .
\label{integral}
\end{equation}
Here $d\Sigma^b$ is defined with an outward pointing normal
which, on the horizon, is
anti-parallel to
the null generator $\chi^b$. Hence the integrand of the second flux
integral is everywhere negative or
vanishing, as long as the stress tensor satisfies
the null energy condition (\ref{null}). Thus
combining eqs.~(\ref{eqone}) and (\ref{integral}) with the
First Law, one has
\begin{equation}
\Delta S=-{2\pi\over\kappa}\int_HT_{ab}\chi^ad\Sigma^b\ge0\ \ .
\label{entropic}
\end{equation}

The details of the gravity theory, and also the precise functional form of
$S$, are irrelevant to establish this result
for quasi-stationary processes. The key requirements were: the First
Law relating asymptotic variations to variations of the horizon
geometry, and the null energy condition to be satisfied by the
matter stress-energy. We emphasize that this
stress-energy tensor includes only contributions from the matter fields,
and not contributions from any higher curvature interactions, as
were considered in the introduction.
We have also made an implicit assumption that the black hole
solution is stable --- {\it i.e.,} the small perturbation introduced
by matter falling in from infinity remains small (and does not
lead to any unstable growing or oscillatory
excitations) so that the system
simple settles down to a new black hole.

We fully expect that the above line of reasoning can be
generalized to cover arbitrary quasi-stationary processes.
For example, the result should hold for the case in which
(possibly charged) matter and
electromagnetic radiation fall into an electrically and/or magnetically
charged black hole. Indeed, we have carried out the analysis
for the case of charged matter (but no radiation) falling onto
a charged black hole. This case is somewhat more complicated
than the vacuum case above.
If one allows infalling electromagnetic radiation then,
due to the nonlinearity of the electromagnetic field stress tensor,
there are generically cross terms between the background field and the
radiation field, yielding $O(\delta)$ contributions to the stress tensor
and to the metric variation. This appears to require a much more
involved---or a much more clever---analysis.
Also, the above
argument does not directly apply in the situation where a
packet of gravity waves drops into the black hole
although it can probably be adapted to cover positive energy
gravitational perturbations.

\section{$R+\alpha R^2$ Theory}

In this section we establish the constancy of the surface gravity
for stationary solutions and the Second Law for arbitrary
dynamical processes involving black holes in the theory
given by a higher curvature action of the form
\begin{equation}
I_{\scriptscriptstyle 0}
=\int d^D\!x\sqrt{-g}\left[{1\over16\pi G}(R+\alpha R^2)+L_m(\psi,g)
\right]
\label{act0}
\end{equation}
where $L_m$ denotes a conventional Lagrangian for some collection of matter
fields, denoted $\psi$. The matter Lagrangian will also
contain the metric, but we assume that it contains no derivatives of
the metric.

The gravitational field equations arising from the action (\ref{act0})
are
\begin{eqnarray}
R_{ab}-{1\over2}g_{ab}R&=&8\pi G\, T^m_{ab}(\psi,g)+2\alpha\,
(\,\nabla_a\nabla_bR-g_{ab}\nabla^2R
\nonumber\\
&&\qquad\qquad-\ R\,R_{ab}+{1\over4}g_{ab}\,R^2\,)\ \ .
\label{movie}
\end{eqnarray}
We will assume that
matter stress-energy tensor, $T^m_{ab}=-{2\over\sqrt{-g}}{\delta\sqrt{-g}
L_m\over\delta g^{ab}}$, does satisfy
the dominant energy condition (\ref{domin}). However, if one regards
the entire expression on the right hand side of this
equation as the stress-energy tensor in Einstein's equations (\ref{eom}),
it is clear that this total stress-energy does not satisfy
any energy condition because of the higher curvature
contributions ({\it i.e.,} the terms proportional to $\alpha$). Thus, as
discussed in the introduction, Hawking's proof of the area
theorem does not apply here, nor can the proof of the
Zeroth Law for stationary black holes in Einstein
gravity be invoked.

For this theory, the black hole entropy appearing in the First
Law (\ref{first}) can be written \cite{visser,wald2,onS}
\begin{equation}
S={1\over4G}\int_{\cal H} d^{D-2}\!x\, \sqrt{h}\, (1+2\alpha\, R)
\label{entrope}
\end{equation}
where the integral is taken over a space-like cross-section of the horizon,
${\cal H}$.  The above form of the black hole entropy is not strictly
justified unless we know that stationary
black holes of the present theory possess a Killing horizon
with constant surface gravity. We shall show below that, for $\alpha>0$,
the surface gravity of a Killing horizon in this theory is necessarily
constant. Note that a number of ambiguities still arise in Wald's construction,
and so eq.~(\ref{entrope}) is the result after making certain natural
choices in the calculation.
None of these ambiguities have any effect when the Noether
charge is evaluated on a stationary horizon\cite{onS}, and hence
eq.~(\ref{entrope}) may be considered in the analysis of sect.~2 since
there the entropy is only compared between the initial and
final stationary black holes. So one knows that the quantity (\ref{entrope})
will always increase in a quasi-stationary process in which a packet of
the matter is dropped into a black hole (since by assumption,
the matter stress-energy satisfies the dominant energy condition
(\ref{domin}), which implies the null energy condition (\ref{null})).

We now extend this result to a classical entropy increase theorem
for any dynamical process involving black holes in this theory.
Our approach will be the following. First, we show that the present higher
curvature theory is equivalent to Einstein gravity
for a conformally related metric coupled to
an auxiliary scalar field, as well as to the original matter fields.
Second, we argue that the black hole entropy in the higher curvature theory
is identical to that in the conformally related theory. Finally, since
Hawking's area theorem holds in the Einstein-plus-scalar theory,
we conclude that the entropy never decreases
in the original theory (\ref{act0}).

The equivalence of the higher curvature theory (\ref{act0}) to Einstein
gravity coupled to an auxiliary scalar field has been discussed previously
by many authors\cite{conform}. The first step
is to introduce a new scalar field $\phi$, and a new action,
which is linear in $R$,
\begin{equation}
I_{\scriptscriptstyle 1}
=\int d^D\!x\sqrt{-g}\left\{{1\over16\pi G}\left[(1+2\alpha\phi)R
-\alpha\phi^2\right]+L_m(\psi,g)\right\}\ \ .
\label{act1}
\end{equation}
The $\phi$ equation of motion is simply $\phi= R$, and
one recovers the original action upon substituting this equation into
eq.~(\ref{act1}) --- {\it i.e.,} $I_{\scriptscriptstyle 1}(\phi= R)
=I_{\scriptscriptstyle 0}$.
In the form of eq.~(\ref{act1}), the action contains no terms that
are more than quadratic in derivatives. This action contains
an unconventional interaction, $\phi R$, however.
Hence in the metric equations of motion,
\begin{eqnarray*}
R_{ab}-{1\over2}g_{ab}\,R&=&8\pi G\, T^m_{ab}(\psi,g)+2\alpha\,
(\,\nabla_a\nabla_b\phi-g_{ab}\nabla^2\phi\\
&&\qquad\qquad-\ \phi\,R_{ab}+{1\over2}g_{ab}\,\phi R
-{1\over4}g_{ab}\,\phi^2\,)\ ,
\end{eqnarray*}
the total stress-energy tensor appearing on the right hand side
still contains some problematic contributions ({\it e.g.,}
$\nabla_a\nabla_b\phi$),
which prevent the dominant or even the null energy conditions
from being satisfied.

The $\phi R$ interaction can be removed by performing the
following conformal transformation
\begin{equation}
{g}_{ab} = (1+2\alpha\phi)^{-{2\over D-2}}\bar{g}_{ab}\ \ .
\label{two}
\end{equation}
In terms of $\bar{g}_{ab}$, the action (\ref{act1}) becomes
\begin{eqnarray}
I_{\scriptscriptstyle 2}
&=&\int d^D\!x\sqrt{-\bar{g}}\left\{{1\over16\pi G}\left[\bar{R}
-{D-1\over D-2}\left({2\alpha\over1+2\alpha\phi}\right)^2\bar{\nabla}_a\phi
\bar{\nabla}^a\phi-\alpha(1+2\alpha\phi)^{-{D\over D-2}}\phi^2\right]\right.
\nonumber\\
&&\qquad\qquad\qquad\qquad
\left.\vphantom{\left[\left({2\alpha\over1+2\alpha\phi}\right)^2\right]}
+(1+2\alpha\phi)^{-{D\over D-2}}L_m(\psi,(1+2\alpha\phi)^{-{2\over
D-2}}\bar{g})
\right\}\ ,
\label{act2}
\end{eqnarray}
which includes the standard Einstein-Hilbert action for $\bar{g}_{ab}$
and the auxiliary scalar $\phi$ with less conventional
couplings --- see below. The $\bar{g}_{ab}$ equations of motion are now
\begin{eqnarray}
\bar{R}_{ab}-{1\over2}\bar{g}_{ab}\,\bar{R}&=&{8\pi G\over1+2\alpha\phi}
T^m_{ab}(\psi,(1+2\alpha\phi)^{-{2\over D-2}}\bar{g})+
{D-1\over D-2}\left({2\alpha\over1+2\alpha\phi}\right)^2
\bar{\nabla}_a\phi\bar{\nabla}_b\phi
\nonumber\\
&&\qquad-\ {1\over2}\bar{g}_{ab}
\left[{D-1\over D-2}\left({2\alpha\over1+2\alpha\phi}
\right)^2\bar{\nabla}_c\phi
\bar{\nabla}^c\phi+\alpha(1+2\alpha\phi)^{-{D\over D-2}}\phi^2\right]\ \ .
\label{moving}
\end{eqnarray}
The most important feature of this final theory for our purposes
is that, assuming $1+2\alpha\phi>0$,
the total stress-energy tensor appearing on the right hand
side above satisfies the dominant (and hence the null)
energy condition for positive $\alpha $ if the original matter Lagrangian
satisfies this energy condition.

Now suppose we have a stationary black hole solution to
(\ref{moving}) whose event horizon is a Killing horizon.
Then, since the dominant energy condition holds
for $\alpha > 0$, the surface gravity must be constant---{\it i.e.,}
the Zeroth Law holds\cite{barcarhaw}. We can therefore identify the entropy
via any of the usual methods {\it e.g.,} Wald's derivation of the First Law.

Given the absence of higher derivative or unconventional gravity
couplings, the black hole entropy for $I_2$
is given by $\bar{S}=\bar{A}/(4G)$,
just as for Einstein gravity. Since the equations of motion (\ref{moving})
are Einstein's equations, and the null energy condition
is satisfied by the total stress-energy tensor,
Hawking's proof of the area theorem is valid for the $I_2$ theory
with the assumptions that cosmic censorship holds
for $\bar{g}_{ab}$  and that $1+2\alpha\phi>0$.
(The latter assumption will be further discussed below.)
Hence there is a classical entropy increase theorem for
the theory defined by $I_2$ in eq. (\ref{act2}).

Now eq.~(\ref{two}) along with $\phi=R$ provides a mapping between
the solutions for the Einstein-plus-scalar theory defined by
$I_{\scriptscriptstyle 2}$,
and the original higher curvature theory defined by
$I_{\scriptscriptstyle 0}$,
in which the metrics are related by a conformal transformation
\begin{equation}
\bar{g}_{ab} = (1+2\alpha R)^{2\over D-2}g_{ab}\ \ .
\label{map}
\end{equation}
The conformal transformation (\ref{map})
preserves the causal structure of the solutions and,
if $g_{ab}$ is asymptotically flat, then so is
$\bar{g}_{ab}$. Thus, if $g_{ab}$ is an asymptotically flat black hole,
then so is $\bar{g}_{ab}$, and they have the same horizon and
surface gravities \cite{tjgk}.
In particular, stationary black hole solutions of the
$I_{\scriptscriptstyle 0}$
theory have constant surface gravity, provided they have Killing
horizons. (Note that, since a Killing vector remains a Killing vector
under
{\it stationary} conformal transformations,
the event horizon of a stationary
$I_{\scriptscriptstyle 0}$ black hole is a Killing horizon if and only if
the same is true for the corresponding $I_{\scriptscriptstyle 2}$ black hole.)
The Zeroth Law therefore holds for $I_{\scriptscriptstyle 0}$.

On the other hand, since the asymptotic
forms of $g_{ab}$ and $\bar{g}_{ab}$ agree, the mass and angular
momenta of the two spacetimes agree.
Also the angular velocities agree, since the null combination
of time translation and rotation Killing fields agree on the
horizon. In short, we have shown all of the ingredients,
other than the entropy, in the
First Law (\ref{first}) agree. Thus, for all variations,
the changes in the entropies must also agree. Therefore the
entropies themselves are equal to within a constant in each
connected class of stationary black hole solutions.
Since the area increase theorem for the Einstein-plus-scalar theory
gives $\delta\bar{S}\ge0$ in any dynamical process connecting two
stationary states, we conclude that $\delta{S}\ge0$ for the corresponding
process in the higher curvature theory.
We have thus established a classical Second Law in the higher curvature
theory defined by the action $I_0$ in eq. (\ref{act0}).

It should be emphasized that the First Law, which applies
to variations away from a stationary black hole background,
does not uniquely determine the form of the entropy for
nonstationary states\cite{wald1,onS}.
In Einstein gravity, since the entropy is proportional to
the horizon area, it has a natural extension to a cross section
of an arbitrary nonstationary black hole horizon, and the area theorem
shows that this nonstationary entropy never decreases, even
{\it during} a dynamical process.
We can obtain a similar result for the higher curvature theory
as follows.

Using the conformal relation (\ref{map}) between the two metrics, the
``barred" entropy (area) can be expressed directly in terms of
$g_{\mu\nu}$:
\begin{equation}
\bar{S}(\bar{g})={1\over4G}\int_{\bar{{\cal H}}}d^{D-2}\!x\,\sqrt{\bar{h}}
={1\over4G}\int_{\bar{\cal H}}d^{D-2}\!x\,\sqrt{h}\,(1+2\alpha\,R)\ \ .
\label{mapent}
\end{equation}
For stationary black holes the right hand side agrees with the
entropy in the higher curvature theory (\ref{entrope}) as determined
directly from the First Law in that theory
(it was already argued above that
$\bar{\cal H}$ corresponds to a cross-section
of the event horizon for the metric $g_{ab}$ as well).
This agreement is explained by the reasoning given
two paragraphs above. In presenting the result (\ref{entrope})
for the black hole entropy as determined by the First Law, we
chose the simplest geometric formula which naturally extends to
a dynamical horizon. Here we have shown that,
by virtue of the area theorem in the conformally related theory,
the entropy given by that particular
formula, reproduced in eq.~(\ref{mapent}),
obeys the Second Law even {\it during} a dynamical process.

The relation (\ref{map}) gives an unambiguous result for the
dynamical entropy and so can be used to resolve
the ambiguities \cite{onS,wald2} inherent in the
Noether charge construction of \cite{wald1}.
In the present case, the alternate proposal
for dynamical entropy of ref.~\cite{wald2},
which used a boost invariant projection,
gives a result for the dynamical entropy that differs from (\ref{mapent})
for non-stationary black holes.
Unless there are two different entropy functionals obeying
a local increase law, it appears that the proposal of
ref.~\cite{wald2} is inconsistent
with the Second Law during dynamical processes in the present theories.

There is one considerable assumption in preceding discussion, which
we have not yet addressed. For the mapping between the
solutions of the two theories (\ref{map}) to exist
and for the total stress energy tensor in eq.~(\ref{moving}) to
satisfy the dominant energy condition (\ref{domin}),
it is necessary that the factor $1+2\alpha R$ is positive.
Thus, given a black hole solution of the higher curvature
theory, one must have $R>-{1\over2\alpha}$ (for positive $\alpha$)
everywhere outside of the black hole and on the event horizon.

{}From the point of view of the Einstein-plus-scalar theory,
one requires that $\phi>-{1\over2\alpha}$ everywhere outside of the event
horizon for the mapping to a solution of the higher curvature
theory to exist. Recall that cosmic censorship was assumed in the proof
of the area increase theorem for this theory. This assumption
rules out dynamical processes in which a black hole begins with
a configuration with $\phi>-{1\over2\alpha}$ everywhere initially, and
evolves to one with $\phi\le-{1\over2\alpha}$ in some region outside
of the horizon. The reason is that, by the equations of motion
(\ref{moving}) when $\phi=-{1\over2\alpha}$, the stress-energy tensor
is singular and hence the Einstein tensor,
$\bar{R}_{ab}-{1\over2}\bar{g}_{ab}\bar{R}$, is singular\cite{footsing}.
Cosmic censorship
for $\bar{g}_{ab}$
would only allow such curvature
singularities to develop
behind the event horizon, and hence rules out any process in which
$\phi=-{1\over2\alpha}$ is reached outside of the horizon.

An alternative argument showing that it is consistent to make the
restriction $\phi>-{1\over2\alpha}$ can be given by considering the character
of the potential term in the Einstein-plus-scalar theory\cite{andy}.
The non-standard kinetic term for the scalar
field $\phi $ in Eq.(\ref{act2}) can be replaced with an ordinary one
by defining a new scalar field $\varphi :=
\beta^{-1}\ln (1+2\alpha \phi )$, where $\beta =\sqrt{8\pi
G(D-2)/(D-1)}$. In terms of $\varphi$ the action $I_2$ becomes
\begin{equation}
I_{\scriptscriptstyle 3}
=\int d^D\!x \sqrt{-\bar{g}}\left[ \frac{1}{16\pi G}
\bar{R}-\frac{1}{2}\bar{\nabla }_a\varphi \bar{\nabla }^a
\varphi - V(\varphi ) + e^{-\frac{D}{D-2}\beta \varphi }
L_m(\psi ,e^{-\frac{2}{D-2}\beta \varphi }\bar{g})\right],
\label{anotherlabel}
\end{equation}
where $V(\varphi )=\frac{1}{64\pi G\alpha }e^{-\frac{D}{D-2}
\beta \varphi }(e^{\beta \varphi }-1)^2 $.
Now the singular point $\phi =-\frac{1}{2\alpha }$ corresponds to $\varphi
\rightarrow -\infty $. Provided $\alpha>0$, the potential $V(\varphi )$
rises exponentially as $\varphi\rightarrow -\infty$.
The term involving the matter Lagrangian has the same exponential for
a prefactor, and so one may worry that it may undermine the barrier
due to $V(\varphi)$. The kinetic terms for the matter fields
will include at least one inverse metric which will
bring the rate of exponential growth down by a factor of
$\exp(2\beta\varphi/(D-2))$ for these contributions.
We will assume that any matter potential is non-negative ---
which is implied by the dominant energy condition (\ref{domin})
for $T^m_{ab}$ --- so that these terms can only increase the potential
barrier as $\varphi\rightarrow-\infty$.
Thus, as long as the metric and matter fields
do not become singular, the dynamics of $\varphi$
as $\varphi\rightarrow -\infty$ will be dominated by
the potential barrier so $\varphi$ will not run off to $-\infty$.
Therefore, initial data satisfying the bound
$\phi>-{1\over2\alpha}$ will evolve within the bound, as long as the other
fields remain nonsingular.

Note that the argument just given breaks down if $\alpha<0$ since the
potential is then {\it negative} and
exponentially {\it falling} as $\varphi\rightarrow -\infty$.
Hence, the theory appears unstable for negative $\alpha$.
The previous argument for non-decreasing entropy
did not seem to require that $\alpha$ be positive
because the null energy condition is satisfied for any $\alpha $.
It did assume cosmic censorship however which, presumably,
would be violated in the unstable theory with $\alpha<0$.
Note also that we previously used this condition on $\alpha$
in order to establish the Zeroth Law.

Given the above arguments that
$\phi=-{1\over2\alpha}$ is never reached outside
the event horizon for positive $\alpha$,
one may also rule out processes in the higher curvature theory
in which a black hole evolves to reach $R=-{1\over2\alpha}$ somewhere
outside of the event horizon. From a superficial
examination of the higher curvature equations of motion (\ref{movie}),
$R=-{1\over2\alpha}$ does not appear to be singular.
However, there
is no obstruction to mapping the initial part of the evolution to
the Einstein-plus-scalar theory, where it becomes a process
leading up to a naked singularity at the point where $\phi=-{1\over2\alpha}$,
as discussed above. Such processes were ruled out by the assumption
of cosmic censorship though, and hence we have also ruled out the
corresponding evolution in the higher curvature theory.

\section{Actions polynomial  in $R$}

The results of the previous section are easily generalized for higher
curvature theories with actions of the form
\begin{equation}
I_{\scriptscriptstyle 0}=\int d^D\!x\sqrt{-g}\left[{1\over16\pi G}
(R+P(R))+L_m(\psi,g)\right]
\label{act00}
\end{equation}
where $P$ is a polynomial in the Ricci scalar,
$P(R)=\sum_{n=2}a_n R^n$.
Introducing an auxiliary scalar field $\phi$, as in (\ref{act1})
of  the preceeding section, this theory can be re-expressed using a new
action linear in $R$
\begin{equation}
I_{\scriptscriptstyle 1}=\int d^D\!x\sqrt{-g}
\left\{{1\over16\pi G}\left[R+P(\phi)+(R-\phi)P'(\phi)
\right]+L_m(\psi,g)\right\}.
\label{act11}
\end{equation}
Here, the primes denote differentiation
of $P$ with respect to $\phi$ --- {\it i.e.,}
$P'(\phi)=\sum_{n=2}n\,a_n\phi^{n-1}$.
The theory defined by this new action is {\it not} precisely
equivalent to the original theory defined by eq. (\ref{act00}).
Rather the $\phi$ equation of motion yields two classes of solutions:
$i)\ \phi=R$ and $ii)\ \phi=\phi_0$ where $\phi_0$ is a constant
satisfying $P''(\phi_0)=0$. Substituting $(i)$ back into the action
(\ref{act11}) yields (\ref{act00}). Thus these solutions correspond
to solutions of the higher curvature theory, which we wish to study.
Substituting $(ii)$ into eq. (\ref{act11}) yields Einstein gravity
with an effective Newton's constant, $G_{eff}=G/(1+P'(\phi_0))$,
and an effective cosmological constant, $\Lambda_{eff}=
(\phi_0P'(\phi_0)-P(\phi_0))/2(1+P'(\phi_0))$.
Thus the latter solutions are spurious for our purposes since
they do not correspond to solutions of the original higher
curvature theory.
In the present analysis we only consider asymptotically flat
solutions, which would rule out the second class of solutions
because of the presence of an effective cosmological constant.
Even in the case of an accidental degeneracy where $\Lambda_{eff}=0$,
one still knows that the asymptotically flat solutions for action
(\ref{act11}) includes all of those for the original action (\ref{act00}).

As in the preceeding section, the action (\ref{act11}) can be
 transformed to Einstein gravity coupled to an
scalar field via the conformal transformation
\begin{equation}
\bar{g}_{ab} = (1+P'(\phi))^{2\over D-2}g_{ab}
\label{Xtran}
\end{equation}
yielding the action\cite{conform}
\begin{eqnarray}
I_{\scriptscriptstyle 2}
&=&\int d^D\!x\sqrt{-\bar{g}}{1\over16\pi G}\left\{\bar{R}
-{D-1\over D-2}\left({P''(\phi)\over1+ P'(\phi)}\right)^2\bar{\nabla}_a\phi
\bar{\nabla}^a\phi
\right.\label{act22}\\
&&\qquad\left.
\vphantom{\left[\left({P''(\phi)\over1+ P'(\phi)}\right)^2\right]}
+(1+P'(\phi))^{-{D\over D-2}}\left[(P(\phi)-\phi P'(\phi))
+16\pi G\,L_m(\psi,(1+P'(\phi))^{-{2\over D-2}}
\bar{g})\right]\right\}.
\nonumber
\end{eqnarray}
Equivalence of (\ref{act11}) and (\ref{act22}) holds provided the conformal
transformation is nonsingular --- {\it i.e.,} $(1+P')>0$.

If the matter Lagrangian $L_m$ yields
a stress-energy tensor satisfying the
dominant energy condition (\ref{domin}),
 then the action $I_2$ also yields a stress-energy tensor
satisfying the dominant energy condition provided
$1+P'(\phi)>0$ and $\phi P'(\phi ) - P(\phi ) > 0 $.
(These conditions on $P$ will be discussed further below.)
In this case, it follows that the surface gravity is
constant over the Killing horizon of
a stationary black hole for $I_2$.
Further, with the  assumption that
cosmic censorship applies in this theory,
one can prove the area
increase theorem, and so then one has shown that $\delta\bar{S}\ge0$
in any evolution of a black hole.

Assuming we have a solution in class $(i)$,
the mapping between
the theories takes the form of a conformal transformation
that goes to the identity at infinity. As above, this
means that a black hole in one theory is mapped to a black hole of the
other theory and, further, that the event horizons, surface gravities,
and entropies of the two black hole solutions coincide.
Thus the constancy of the surface gravity and
the area increase theorem
of the Einstein-plus-scalar theory translate to the Zeroth Law and
an entropy increase theorem for the original higher curvature theory,
respectively.

The mapping between the solutions of the two theories yields
a formula for the entropy in the higher curvature theory:
\begin{equation}
\bar{S}(\bar{g})={1\over4G}\int_{\bar{{\cal H}}}d^{D-2}\!x\,\sqrt{\bar{h}}
={1\over4G}\int_{\bar{{\cal H}}}d^{D-2}\!x\,\sqrt{h}\,(1+P'(R))\ \ .
\label{PRentropy}
\end{equation}
As expected one recovers the same expression for the black hole
entropy in the higher curvature theory that was determined
by directly examining the First Law in that theory for variations
from stationary black holes\cite{onS}.
Note that the conformal transformation yields an unambiguous
definition of the dynamical (non-stationary) black hole entropy for the higher
curvature theory, whereas the entropy functional determined from the First Law
is not unique\cite{wald1,onS}.

Of course, the equivalence of the dynamics defined by the actions
$I_{\scriptscriptstyle 0}$ and
$I_{\scriptscriptstyle 2}$  required that $\phi=R$, and
that $1+P'(R)>0$ everywhere for solutions of the higher curvature theory.
The latter assumption requires that $R$ lie within some
domain including zero,
whose precise boundaries will be defined by the couplings $a_n$
appearing in $P(R)$.
Alternatively, there are restrictions on allowed values
of the auxiliary scalar $\phi$ in the Einstein-plus-scalar theory.
For our purposes it is enough that these restrictions hold
outside the horizon and on an open set including the horizon.

One can argue as in the preceding section that cosmic censorship
rules out processes in which a black hole evolves from a configuration
with $1+P'>0$ to one in which this inequality is violated.
This is done by rewriting the action yet one more time,
as in the preceding section,  with the change of variables $\varphi :=
\beta^{-1}\ln (1+P'(\phi) )$, where $\beta =\sqrt{8\pi
G(D-2)/(D-1)}$. In terms of $\varphi$ the action $I_2$ becomes
\[
I_{\scriptscriptstyle 3}=\int d^D\!x \sqrt{-\bar{g}}\left[\frac{1}{16\pi G}
\bar{R}-\frac{1}{2}\bar{\nabla }_a\varphi \bar{\nabla }^a
\varphi - V(\varphi ) + e^{-\frac{D}{D-2}\beta \varphi }
L_m(\psi ,e^{-\frac{2}{D-2}\beta \varphi }\bar{g})\right],
\]
where now
$V(\varphi )=\frac{1}{16\pi G}e^{-\frac{D}{D-2}
\beta \varphi }(\phi P'-P) $.
Equivalence of $I_2$ and $I_3$ requires
that one can invert $1+P'(\phi)=e^{\beta\varphi}$ for $\phi=\phi(\varphi)$.
This requires that $P''$ has a definite sign in the domain
of interest for $\phi$ (which includes $\phi=0$).

As before, the singular point $1+P'(\phi)=0$ corresponds to $\varphi
\rightarrow -\infty $. There is a potential barrier as
$\varphi\rightarrow -\infty $ provided
$\phi P'-P$ is positive. Note that $\phi P'-P$ vanishes at
$\phi=0$, and $(\phi P'-P)'=\phi P''$. If we restrict $P''$
to be positive, $(\phi P'-P)$ will be positive for all
$\phi\ne 0$.
Thus we can argue, as in the preceding section, that if
$P''>0$ (and the matter
stress energy satisfies the dominant energy condition), initial data
satisfying the bound $1+P'>0$ outside the horizon will evolve within
this bound, as long as the other fields remain nonsingular outside
the horizon. As a bonus, $P''>0$ implies the positivity of
$\phi P'-P$, which was required for the dominant energy condition
(and therefore the proof of the Zeroth Law) to hold. Actually,
given that $P''>0$, $1+P'$ (which is one at $\phi=0$) will generically
reach zero at some negative value of $\phi$, which then defines the
boundary for the range of interest.
It is enough then to require $P(\phi)''>0$ for
$\phi\ge \phi_1$, where $\phi_1$ denotes the negative
value of $\phi$ nearest the origin for which
$1+P'$ vanishes. If $P''<0$ in the range of interest,
the theory is probably unstable, as in the
$R+\alpha R^2$ theory with negative $\alpha$.

\section{Direct proof of the Second Law}

Our method of establishing the Second Law for certain higher curvature
theories used the fact that these theories are conformally
related to ordinary Einstein theories in which the area theorem holds.
This special feature of these theories is not shared by most
higher curvature theories, so it would be
interesting to see how the Second Law could be established {\it directly}
in these theories, without making
use of the conformal transformation technique.
Such an exercise would be instructive for efforts to establish entropy
increase theorems for theories that are not susceptible to the
conformal transformation ``trick".
In order to gain some insight into this question,
in the present section we will construct such a direct proof.

Suppose that the black hole entropy of a gravity theory takes
the following form
\begin{equation}
S = \frac{1}{4G}\int_{\cal H}d^{D-2}\!x\, \sqrt{h}\, e^\rho,
\label{expent}
\end{equation}
where $e^\rho$ is a scalar function of the local geometry at the horizon.
For the class of theories considered in the preceding section  one has
$e^\rho=1+P'(R)$. The method to be used here will rely critically on
the fact that $e^\rho$ is necessarily positive, and $\rho=0$
when the curvature vanishes.

We wish to consider the change of this entropy along the null
congruence generating the event horizon under any dynamical evolution.
Let $k^a$ be the null tangent vector field of the horizon generators with
respect to the affine parameter $\lambda $.
Then one has
\[
S'=\frac{1}{4G}\int_{\cal H}d^{D-2}\!x \sqrt{h}\, e^\rho\, \tilde{\theta}
\]
with
\begin{equation}
\tilde{\theta}:= \theta +\partial_\lambda\rho,
\label{defcon}
\end{equation}
where $\theta=d(\ln\sqrt{h})/d\lambda =\nabla_ak^a$
is the expansion of the horizon generators.

Now the question
is whether or not there can exist a point along the null geodesics at
which $\tilde{\theta }$ becomes negative. In order to answer this
question, we use the Raychaudhuri equation, as in the proof of area theorem,
to obtain an expression for $\partial_\lambda\tilde{\theta}$:
\begin{eqnarray}
\partial_\lambda\tilde{\theta}&=& \partial_\lambda\theta
+\partial_\lambda^2\rho
\nonumber     \\
&=& -\textstyle{\frac{1}{D-2}}{\theta }^{2}-
{\sigma }^2 - k^ak^bR_{ab} + k^ak^b\nabla_a\nabla_b\rho,
\label{newray}
\end{eqnarray}
where $\sigma^2$ is the  square of the shear.

For the $R+P(R)$ theories, it is easy to see that the equations of
motion imply that
$k^ak^b(R_{ab}-\nabla_a\nabla_b\rho)=(8\pi G\,e^{-\rho}T^m_{ab}+
\nabla_a\rho\nabla_b\rho)k^ak^b$,
which is non-negative provided the null energy condition holds for the
matter fields (and $e^\rho>0$). Thus in those theories one has
\[
\partial_\lambda\tilde{\theta }
\leq -\textstyle{\frac{1}{D-2}}{\theta }^2,
\]
or
\begin{equation}
\partial_\lambda[\tilde{\theta}^{-1}]
\ge\frac{1}{D-2}(\theta/\tilde{\theta})^2.
\label{ineq}
\end{equation}

Now we follow Hawking's proof of the area theorem, with
$\tilde{\theta}$ in place of $\theta$.
Suppose at some point on the horizon we have $\tilde{\theta}<0$.
Then in a neighborhood of that point one can deform a space-like slice of
the horizon slightly outward to obtain a compact space-like surface
$\Sigma$ so that $\tilde{\theta}<0$ everywhere on $\Sigma$,
$\tilde{\theta}$ being
defined along the outgoing null geodesic congruence orthogonal to $\Sigma$.
If cosmic censorship is assumed, then there is necessarily some
null geodesic orthogonal to $\Sigma$ that remains on the boundary
of the future of $\Sigma$ all the way out to ${\cal I}^+$\cite{areathm}.
Asymptotic flatness (where components of the Riemann tensor in an
orthonormal frame all fall off at least as $r^{1-D}$)
implies that $\rho\rightarrow 0$ like $\lambda^{1-D}$
at infinity, whereas $\theta$ goes like $\lambda^{-1}$, where $\lambda$
is the affine
parameter along an outgoing null geodesic. Therefore
$\theta/\tilde{\theta}\rightarrow 1+O(\lambda^{(1-D)})$,
so the inequality (\ref{ineq}) implies that, as one follows the geodesic
outwards from $\Sigma$, $\tilde{\theta}$ reaches $-\infty$
at some finite affine parameter.
Since $\tilde{\theta}=\theta+\partial_\lambda\rho$, this means that either
$\theta$  or $\partial_\lambda\rho$ goes to  $-\infty$. In the former case we
have a
contradiction, as in the area theorem, since it implies there is a
 conjugate point on the geodesic, which cannot happen since the
geodesic stays on the boundary of the future of $\Sigma$ all the way out to
${\cal I}^+$. In the latter case we have a naked singularity,
since $\partial_\lambda\rho=e^{-\rho}
\partial_\lambda e^\rho$, and $e^\rho$ was assumed from
the beginning in (\ref{expent}) to be a nonvanishing function
of the curvature.

For the $P(R)$ theories we have in particular
$\partial_\lambda\rho=k^a\nabla_a \rho=(1+P')^{-1}P''k^a\nabla_aR$.
In the preceding section we argued that $1+P'$ never goes to zero outside
the horizon in the case of a theory with $P''>0$, so for these theories
the divergence of $\partial_\lambda\rho$ implies divergence of
$R$ or $k^a\nabla_a R$,
but these divergences also violate cosmic censorship.
Therefore we conclude that cosmic censorship and the null
energy condition for
the matter imply that the black hole entropy (\ref{expent})
can never decrease for the stable theories.
Note that in making this argument we have used the condition
$1+P'>0$ that was established via the conformal transformation
trick, so we do not really have a fully ``direct proof" of the
Second Law.

The above argument suggests that the ``weakest" naked singularity which
creates a violation of the Second Law would be a divergence in $k^a\nabla_a R$.
It seems that this could happen even if the curvature itself is nonsingular
everywhere. However, if one imposes also the equations of motion of the
theory, then a divergence in $k^a\nabla_a R$ would necessarily entail also
a divergence in either the curvature tensor or the matter stress tensor.

As a final note, we demonstrate that the ``extended" Raychaudhuri
equation (\ref{newray}) lends itself to a
``physical process" derivation of the First Law\cite{waldbook2},
and also of the form of the entropy,
at least for the actions polynomial in the Ricci scalar.
A sketch of such a derivation follows:
Following the quasistationary process discussion in section 2,
one concludes that
the First Law is satisfied if there is an entropy functional $S$
satisfying the equality in (\ref{entropic}),
\[
{\kappa\over2\pi}\Delta S=
\Delta M-\Omega^{\scriptscriptstyle(\alpha)}
\Delta J_{\scriptscriptstyle (\alpha)}=
-\int_HT_{ab}\chi^ad\Sigma^b\ \ .
\]
Now on the horizon, where $\chi^a$ is null,
the equations of motion imply that $8\pi G T_{ab}\,\chi^a\chi^b=
e^\rho \chi^a\chi^b(R_{ab}-\nabla_a\nabla_b\rho-\nabla_a\rho\nabla_b\rho)$
where $e^\rho=1+P'(R)$. The horizon generating Killing field $\chi^a$ is
related to the affinely parameterized null tangent to the
horizon $k^a\partial_a=
\partial_\lambda$ by $\chi^a\partial_a=\kappa\lambda
k^a\partial_a$\ \cite{waldbook2}. Further the volume element in the
above flux integral may be written: $d\Sigma^a=-d^{D-2}x\sqrt{h}\,
d\lambda\,k^a$ \cite{waldbook2}.
Hence using the ``extended'' Raychaudhuri equation
(\ref{newray}) and the equations of motion,
and neglecting terms of higher than linear order in the perturbation,
the above equation yields
\begin{eqnarray*}
{\kappa\over2\pi}\Delta S&=&
\kappa\int_Hd^{D-2}x\sqrt{h}\,d\lambda\,\lambda k^ak^bT_{ab}\\
&=&{\kappa\over8\pi G}\int_Hd^{D-2}x\sqrt{h}\,d\lambda\,\lambda\,e^\rho
\,k^ak^b(R_{ab}-\nabla_a\nabla_b\rho)\\
&=&-{\kappa\over8\pi G}\int_Hd^{D-2}x\sqrt{h}\,d\lambda\,\lambda\,e^\rho
\partial_\lambda\tilde{\theta}\\
&=&-{\kappa\over8\pi G}\oint d^{D-2}x\sqrt{h}\lambda e^\rho\tilde{\theta}
\biggr|^{\Sigma_f}_{\Sigma_i}+{\kappa\over8\pi G}\int_Hd\lambda\,
d^{D-2}x\sqrt{h}\,e^\rho\tilde{\theta}\ \ .
\end{eqnarray*}
In this final expression, $\tilde{\theta}$
vanishes on the final and initial horizon slices where the horizon
is stationary, and so the first contribution is zero. By
the definition of $\tilde{\theta}$ in eq.~(\ref{defcon}), we see
that
$\sqrt{h}\,e^\rho\tilde{\theta}$ is the total derivative
$\partial_{\lambda}(\sqrt{h}\,e^\rho)$, so the second
contribution is just ${\kappa\over2\pi}\Delta S$
where $S$ is precisely the entropy given in eq.~(\ref{PRentropy}).
The form of the entropy functional for these higher curvature theories
can thus be inferred directly by consideration of quasistationary
accretion processes.

\section{Discussion}

In this paper, we have presented two cases where a classical entropy
increase theorem applies in higher curvature gravity. These are:
\begin{itemize}
\item
For quasi-stationary processes in which a (vacuum) black
hole accretes positive energy matter --- {\it i.e.,} $T^m_{ab}\ell^a\ell^b\ge0$
for any null vector $\ell^a$ --- the Second Law is a direct consequence
of the First Law of black hole mechanics, independent
of the details of the gravitational action.
\item
For higher curvature theories of the form (\ref{act00})
the black hole entropy is given by
\begin{equation}
S(g)={1\over4G}\int_{{\cal H}}d^{D-2}\!x\,\sqrt{h}\,(1+P'(R))\ \ .
\label{noise}
\end{equation}
This entropy satisfies
the Second Law in any processes involving matter fields
that satisfy the null energy condition.
Our proof of the Second Law requires that the coupling constants
$a_n$ appearing in $P(R)$ be restricted in such a way that
$P''(R)$ is positive for positive $R$,
and also between $R=0$ and the largest
negative value of $R$ where $1+P'(R)$ vanishes.
The latter ensures that $1+P'(R)$ is positive everywhere
outside and on the event horizon of the black hole spacetimes.
\end{itemize}

The expression $1+P'(R)$ must be positive in order to implement the
conformal transformation between the original higher curvature theory
and the Einstein-plus-scalar theory, and also to ensure the
null energy condition is satisfied in the latter theory.
This positivity
is also an essential ingredient for the direct proof in section 5.
It is interesting that
precisely the same expression plays the role of the entropy
surface density in eq. (\ref{noise}). Thus the positivity restriction
translates on the horizon to the condition that the local entropy
density should be positive everywhere. In particular it requires
that the total black hole entropy is always positive.
The latter is a minimum requirement that must be fulfilled
if this entropy is to have a statistical mechanical origin.
The fact that we actually require
a {\it local} positivity condition on the entropy density
is suggestively consistent with the idea that this density may have a
statistical interpretation.
In any event, these (and other higher curvature) theories
may provide a more refined test of the various proposals to
explain the statistical origin of black hole entropy.

The direct proof of the Second Law (in section 5)
is essentially a translation, via the conformal transformation,
of Hawking's proof of the area theorem applied to the
Einstein-plus-scalar theory. Nevertheless, it provides
an illustration of how one might hope to prove an entropy
increase theorem for other higher curvature theories.
Naively, with the assumption of cosmic censorship,
this proof can be extended to theories with interactions
of the form $R^{2n+1}\nabla^2 R$. A closer examination of the latter
theories indicates that they are unstable however, and so the assumption
of cosmic censorship appears unlikely to hold. This highlights
the problem that, in dealing with the higher curvature theories directly,
establishing the stability of asymptotically flat solutions requires
an involved analysis.
In fact, even for our direct proof in section 5, we relied on
results about the stability of the theories
derived in sections 3 and 4 by examining the Einstein-plus-scalar theory.

An obstacle to constructing a direct proof of the Second Law in general
is that the entropy as determined from the First Law
does not uniquely determine the form of the dynamical entropy.
Thus, to begin, one would not know for which entropy
density one should be attempting to prove an increase theorem.
In the higher curvature theories considered in sections 3 and 4,
this ambiguity is resolved by the conformal transformation, which
yields precisely eq. (\ref{noise}) when inserted into $\bar{S}=\bar{A}/(4G)$.
In ref.~\cite{onS}, the present authors introduced an alternative
construction for black hole entropy involving field redefinitions,
which also appears to avoid any ambiguities. The form of the higher
curvature actions for which this approach is applicable is not
completely general, but it does extend beyond those theories
considered in this paper.

In Einstein gravity, the Zeroth Law for a Killing horizon can be proved if the
dominant energy condition is assumed\cite{barcarhaw}.
Through the conformal transformation technique,
this proof was extended to the higher curvature theories
introduced in sections 3 and 4, at least with certain restrictions
on the coupling constants.
In the case of a regular bifurcate Killing horizon,
one can show that the surface gravity is constant
irrespective of the underlying gravitational
dynamics\cite{raczwald}.
However, there is no independent proof that  Killing horizons
in a general theory necessarily possess a regular bifurcation surface,
and so for general higher curvature theories the validity of the
Zeroth Law remains an important open question.

It is worth emphasizing that unless the event horizon
is a Killing horizon,
the abovementioned proofs of the Zeroth Law are not applicable.
In Einstein gravity, Hawking proved that the
event horizon of a stationary black hole must be a Killing
horizon\cite{areathm}.
To our knowledge, this proof has not been extended to general higher
curvature theories, or even to higher dimensional Einstein gravity.
For the higher curvature theories
considered in this paper,
at least in four dimensions, it seems likely that
Hawking's proof can be imported via the conformal transformation relating
the theory to Einstein gravity with matter.
For stationary black
holes in more general higher curvature theories,
whether or not stationary event horizons are necessarily Killing
horizons is another important open question.

We now come to considering the two shortcomings of the calculations
presented in this paper. For all of the cases considered here,
the dominant energy condition must hold for the matter fields.
Within the present framework, though, it is natural
to include higher derivative matter couplings
({\it e.g.,} $R(\nabla\phi)^2$ or $(\nabla^2\phi)^2$) as well as
higher curvature interactions. Generally the former
will spoil this positive energy condition. This is perhaps less
a criticism of the discussion of actions polynomial in the Ricci
scalar since they are already theories with a restricted set of
interactions.

The second short-coming is revealed by the instabilities faced in
sections 3 and 4. There our proof of the Second Law failed for certain
values of the coupling constants
({\it e.g.,} $\alpha<0$ for the $R^2$ theory) because we found
these theories to be unstable, and hence it appeared that
the conformal transformation technique could not be implemented and
cosmic censorship was not a valid assumption.
Considering the $R^2$ theory represented that the action (\ref{anotherlabel})
with the auxiliary scalar $\varphi$, we see that the scalar potential
is proportional to $1/\alpha$. Thus the instability is nonperturbative
in the higher curvature coupling constant.
Further, a perturbative analysis (around flat space) reveals
unstable modes with imaginary frequencies of the order of $1/\alpha$.
We expect these remarks also to be true of the unstable theories in section
4. The original framework, which we set out for our investigations, though,
was Einstein gravity {\it perturbatively} corrected by higher curvature
corrections. The problem with the present analysis is that we
are actually taking these theories at their face value, rather
than treating the higher curvature terms perturbatively.
This suggests that our analysis should be modified to incorporate
the ideas of perturbative reduction\cite{redux}.
One might hope that the problems with instability
and cosmic censorship would be avoided in this way.
Since the perturbative treatment would extend
to all of the higher derivative interactions, including those of the
matter fields, such an approach may also be able to circumvent the
requirement that the full matter stress energy tensor satisfy the
null energy condition\cite{footnn}.

Establishing the Second Law for higher curvature theories within
a perturbative framework would be a valuable extension of our
present results, since we expect  nonperturbative instabilities
to be a generic feature of these theories when they are considered
as fundamental\cite{wood}. If one were to find that no Second Law holds
{\it even perturbatively}
for certain interactions or certain values of the coupling constants,
one might suspect that those effective actions are unphysical.
Perhaps the requirement that the entropy (or even the entropy density)
be positive might provide a further restriction on the form of
physically relevant effective actions.

In this paper we have only considered an intrinsic or classical Second
Law --- {\it i.e.,} we have only dealt with the increase of the black
hole entropy alone. In general relativity, we know
that the effective transfer of negative energy from quantum fields to
a black hole can lead to a decrease in the horizon entropy
({\it i.e.,} horizon area), and the same is true
for these higher curvature effective theories since
black holes still produce Hawking radiation in these theories.
Thus it is important to ask whether a generalized
Second Law ($\delta (S_{BH}+S_{outside})\ge 0$) holds.
In general relativity, there are arguments that the
generalized Second Law applies for quasi-stationary
processes involving positive energy matter\cite{gsl}.
These arguments seem to carry over to stable
higher curvature gravity theories as well,
since they do not involve the equations of motion but rather lean on the
First Law and the maximum entropy property of thermal radiation.

Another approach to this question would be to incorporate
the effects of the Hawking radiation in the effective
action by the introduction of nonlocal
contributions\cite{nonlocal,nonlocal2}.
In two-dimensions where an explicit nonlocal term can be
calculated\cite{nonlocal2}, Wald's techniques have been
applied to determine the nonlocal (radiation) contribution to the
geometric horizon entropy\cite{entwo}. Further, in that
model, one can show that the generalized entropy,
which now includes the contributions of the Hawking
radiation, will satisfy a Second Law, even for evaporating
black holes.\cite{fiola} Some model independent constructions for the
nonlocal action exist in four dimensions\cite{nonlocal},
and so one could in principle apply Wald's techniques
to develop an expression for the black hole entropy
in these theories. Perhaps
the requirement that this entropy satisfies the Second Law
would impose useful restrictions on the underlying nonlocal action.
In any event, addressing the validity of the generalized
Second Law remains an important open problem.

\vskip 1cm
We would like to acknowledge useful discussions with R. Woodard
and J. Friedman. R.C.M.\ was supported by NSERC of Canada, and Fonds FCAR du
Qu\'{e}bec.  T.J.\ and G.K.\ were supported in part by NSF Grant~PHY91--12240.
We would also like to thank the the ITP, UCSB for their
hospitality during the early stages of this work.
Research at the ITP, UCSB was supported by NSF Grant~PHY89--04035.
T.J.\ would like to thank the Institute for Theoretical Physics
at the University of Bern for hospitality and Tomalla Foundation
Zurich for support while some of this work was done.


\begin{thebibliography}{99}
\bibitem
{footsing}
We assume that the singularities in the individual terms in the
stress-energy tensor of eq.~(\ref{moving}) do not cancel to
avoid a curvature singularity.
\bibitem
{footnn}
The starting point for such a perturbative analysis is a solution
of Einstein's equations. A vacuum solution with $R=0$ would
therefore be unaffected perturbatively
for the theories considered in sections 3 and 4. It is only
in the presence of matter that $R\ne0$, and hence that the special class of
higher curvature terms considered there
would play a role in modifying the solutions
and the black hole entropy.
\end{thebibliography}
\end{document}